\documentclass[11pt]{article}
\usepackage{jheppub}
\usepackage{mymacros}

\title{Relational dynamics in axion-de Sitter universes}
\author[a,b]{Sergio E. Aguilar-Gutierrez}
\affiliation[a]{Qubits and Spacetime Unit, Okinawa Institute of Science and Technology Graduate University, 1919-1 Tancha, Onna, Okinawa 904 0495, Japan}
\affiliation[a]{Institute for Theoretical Physics, KU Leuven,\\ Celestijnenlaan 200D, B-3001 Leuven, Belgium}
\emailAdd{sergio.ernesto.aguilar@gmail.com}
\abstract{We study the evolution of codimension-one maximal volume and related geometric observables according to a pair of dynamical reference frames (i.e. observers) in asymptotically de Sitter universes in the presence of axion matter. This sources a Euclidean wormhole entangling the two universes, thereby changing the spacetime topology. The system is prepared in a two-copy Hartle-Hawking state by slicing a Euclidean wormhole. We derive the evolution of codimension-one observables anchored to the worldline reference frame in each universe. We investigate how the axion charge competes with the cosmological constant in the evolution of the observables. The maximal volume increases nearly exponentially for low axion charge, while it decreases to a vanishing value for the maximal axion charge allowed by de Sitter space.}

\begin{document}

\maketitle

\section{Introduction}\label{sec:intro}
Cosmological spacetimes, and particularly closed universes, are usually full of enigmatic characteristics in quantum gravity. The arguably best known examples include the problem of time \cite{Isham:1992ms} due to the lack of an external notion of time leading to the Wheeler–DeWitt Hamiltonian constraint \cite{DeWitt:1967yk,Wheeler:1968iap}, and the puzzle of the one-dimensionality of the Hilbert space of closed universes \cite{Almheiri:2019hni,Penington:2019kki,Kirklin:2023yqs,Balasubramanian:2023xyd}, which has been more recently ignited by \cite{Harlow:2025pvj,Abdalla:2025gzn} (see also \cite{Akers:2025ahe,Chen:2025fwp,Nomura:2025whc,Blommaert:2025rgw} for follow-up developments). The role of the sum over topologies in the path integral is at the heart of the one-dimensionality of the Hilbert space of closed universes, which might seem to be in contradiction to the non-trivial evolution with respect to a classical observer. Most of the approaches to this problem rely on identifying diffeomorphism invariant observables respect to a reference frame. Relational dynamics (see e.g. \cite{Krumm:2020fws,Hohn:2017cpr,Hoehn:2019fsy,Hoehn:2020epv,Hoehn:2023ehz,Vanrietvelde:2018pgb,Vanrietvelde:2018dit,delaHamette:2021oex,Hohn:2018toe,Hohn:2018iwn,Hoehn:2021flk,Giacomini:2021gei,Yang_2020,DeVuyst:2024uvd,DeVuyst:2024pop,Witten:2023xze} among many others) is a framework to describe quantum or classical observables that are dressed with respect to dynamical reference frame in a gauge invariant matter. This provides a physically meaningful way to describe observables in closed universes, which is the main premise of our work. 

Due to technical challenges, the role of non-trivial topology in closed universes with a positive cosmological constant remain vastly unexplored in comparison with the negative cosmological constant case \cite{Harlow:2025pvj,Abdalla:2025gzn,Akers:2025ahe} so far. Nevertheless, there has been recent progress in addressing some of these challenges with a specific toy model, Einstein gravity with axion matter content and a positive cosmological constant \cite{Gutperle:2002km,Halliwell:1989pu,Aguilar-Gutierrez:2023ril,Aguilar-Gutierrez:2023hls,Myers:1988sp,klebanov1989wormholes,Blommaert:2025bgd}, which retains useful characteristics to study more realistic cosmological spacetimes. Axion particles result from Kaluza-Klein compactification in string theory, see \cite{Svrcek:2006yi} for a review. This leads to different top-down examples of axion wormholes \cite{Giddings:1987cg,Giddings:1989bq,Loges:2023ypl,Andriolo:2022rxc,Marolf:2021kjc}, although not for the wormholes with a positive cosmological constant involved in our study. A proper understanding of the Euclidean geometry of this solution, and its interpretation in quantum cosmology only recently appeared in \cite{Aguilar-Gutierrez:2023ril}. In Lorentzian-signature, the system we study is a pair of axion-de Sitter (dS) universes represent spatially closed FLRW cosmologies, which are prepared by a two-copy Hartle-Hawking (HH) state \cite{Hartle:1983ai}, and where the axion matter is described by an ultra-stiff fluid, which has an equation of state $\rho=p$, indicating energy density and pressure respectively. The analysis of the time direction with respect to matter inhomogeneities shows that the universes have inverted arrows of time which were interpreted as bouncing cosmologies mediated through the Euclidean wormhole connecting the universes \cite{Aguilar-Gutierrez:2023ril}. Previous work in this model also studied notions of entropy, and late-time two-point correlator functions \cite{Aguilar-Gutierrez:2023hls}.

Regardless of the apparent simplicity in this model, it raises important puzzles. For instance, it was realized in \cite{Halliwell:1989pu} that the saddle points with multiple spheres in the model we consider may result in an unbounded on-shell action. However, the same authors showed that when one treats the metric as complex, there appears a cancellation between saddle point contributions. However, it has been more recently brought to attention \cite{Blommaert:2025bgd} that the type of real saddle points found by \cite{Aguilar-Gutierrez:2023hls} in the same theory as \cite{Halliwell:1989pu} do indeed lead to an unbounded negative on-shell action, which means that the corresponding Euclidean path integral will generally be ill-defined at the saddle point approximation. This means that the Euclidean path integral might not be well approximated by a contour passing through the saddle points representing an arbitrary number of connected universes. For these reasons, it is crucial to understand in better detail what are the properties of the wormhole solutions in \cite{Aguilar-Gutierrez:2023hls}, which might allow us to address the puzzles they lead to.

The observables in this work also take a crucial role in holographic complexity conjectures \cite{Susskind:2014rva,Stanford:2014jda,Couch:2016exn,Brown:2015bva,Brown:2015lvg,Belin:2021bga,Belin:2022xmt}; particularly the complexity=volume \cite{Susskind:2014rva,Stanford:2014jda}, and other generalizations, called complexity=anything \cite{Belin:2021bga,Belin:2022xmt}, which will be evaluated on codimension-one constant mean curvature (CMC) slices \cite{Belin:2022xmt,Jorstad:2023kmq,Aguilar-Gutierrez:2023tic,Aguilar-Gutierrez:2023zqm,Aguilar-Gutierrez:2023pnn}.
These proposals defined as geometric invariant observables, that are conjectured to be holographically dual to some notion of quantum complexity (see recent review in \cite{Chapman:2021jbh,Baiguera:2025dkc,Rabinovici:2025otw,Nandy:2024evd,Chen:2021lnq,Myers:2024vve}). Furthermore, there has been remarkable recent progress in adapting the above proposals beyond asymptotically anti-de Sitter (AdS) space, especially for asymptotically dS space through stretched horizon holography\footnote{See \cite{Galante:2023uyf} for a recent review.} \cite{Susskind:2021esx,Jorstad:2022mls,Auzzi:2023qbm,Anegawa:2023wrk,Baiguera:2023tpt,Anegawa:2023dad,Aguilar-Gutierrez:2023zqm,Aguilar-Gutierrez:2023tic,Aguilar-Gutierrez:2023pnn,Aguilar-Gutierrez:2024rka,Chapman:2021eyy}. However, given that dS/conformal field theory (CFT) holography is not sufficiently developed, we do not attempt to interpret our study of bulk relational observables in terms of complexity. Instead, the aim of our study is to learn about the internal evolution of relational observables effects in a closed universe with non-trivial topology. Nevertheless, assuming that some of these conjectures might hold in more general spacetimes, our work also contributes to the literature by providing an example with a more intricate topology.

Motivated by these developments, we study the implications of anchoring the simplest gravitational observables, such as codimension one volumes, to dynamical reference frames in the same background as \cite{Aguilar-Gutierrez:2023hls}. For technical simplicity in studying gauge invariant observables in closed universes in the semiclassical limit, we follow a similar approach as \cite{Chandrasekaran:2022cip,Witten:2023qsv,Witten:2023xze}, where one considers the $G_N\rightarrow0$ regime and the observables are dressed with respect to dynamical reference frames at the classical level, like in \cite{Goeller:2022rsx}, following a worldline geodesic.\footnote{Alternatively, one might resort to the late time analysis where gravity at $\mathcal{I}^+$ of the asymptotically dS universe is weak, as in the dS/CFT holographic approach \cite{Strominger:2001pn}.} This allows us to characterize relational observables in terms of the axion charge and the dS length scale, which determines the wormhole geometry. The strategy goes as follows. We locate a geodesic worldline observer in each of the universes equipped with a clock, where we assume that they can be synchronized through the HH state preparation. They will measure the maximal codimension-one volume and more general curvature invariant observables as a function of the global time. Since the universe are connected through a wormhole, the extremal volume slices connecting the observers also intercept the Euclidean wormhole in the past of each observer. Our results show that the time evolution of the relational observables generically increases exponentially in terms of the global time of the FLRW cosmologies when the axion charge is dilute enough; while it can also decrease and reach a constant value for a sufficiently dense axionic charge in this background. These effects also depend on the mean curvature of the extremal surfaces where the observables are evaluated.

\textbf{Structure:} In Sec. \ref{sec:review} we provide a lightning review about the axion-dS universes. In Sec. \ref{sec:CV} we study the maximal codimension-one volume anchored with respect to the pair of observers, and we discuss the effects of the axion charge on the evolution of the volume. In Sec \ref{sec:CAny}, we investigate more general codimension-one curvature-invariants observables using CMC slices covariantly define the extremal surfaces, and the resulting modification on the rate of growth as the mean curvature increases. Finally, Sec. \ref{sec:Discussion} includes a summary of the manuscript and some future directions. For the convenience of the reader, we also include App. \ref{app:notation} with a summary of the notation in our work.

\section{Geometry of axion-de Sitter wormholes and universes}\label{sec:review}
In this section, we briefly review the properties of axion-dS universes, which are prepared from an Euclidean wormhole. More geometric details can be found in \cite{Aguilar-Gutierrez:2023ril} as well as some results on the on-shell action of these solutions and their perturbative stability; while a discussion about entanglement and late-time bulk correlators in the Lorenzian theory is given in \cite{Aguilar-Gutierrez:2023hls}.

Our starting point is $D$-dimensional Euclidean Einstein gravity in the presence of axion matter content and a positive cosmological constant:
\begin{equation}\label{eq:onshell fundamental}
I=\int\qty[-\frac{1}{2\kappa_D^2}\star(R-2\Lambda)+\frac{1}{2}\star H_{D-1}\wedge H_{D-1}]~,
\end{equation}
where $H_{D-1}$ is the axion flux field, which is Hodge dual to the axion field $\chi$ (i.e. $H_{D-1}=\star d\chi$); $\kappa_D^2=8\pi G_N$ and we will consider a cosmological constant
\begin{equation}\label{eq:Lambda C.C.}
    \Lambda=\frac{(D-1)(D-2)}{2L^2}>0~.
\end{equation}
The presence of the axion flux field in (\ref{eq:onshell fundamental}) produces Strominger-Giddings wormhole \cite{Giddings:1989bq}-type of solutions. We study the saddle point of the path integral in the minisurperspace approximation while considering spherical symmetric solutions for this theory:
\begin{equation}\label{eq:metric to be gauged}
    \rmd s^2=N(\tau)^2\rmd\tau^2+a(\tau)^2\rmd\Omega_{D-2}^2~,
\end{equation}
where we denote
\begin{equation}\label{eq:metric regular}
    \rmd\Omega_{D-1}^2=\rmd\theta_1^2+\cos^2\theta_1\rmd\theta_2+\dots+\cos^2\theta_1\dots\cos^2\theta_{D-2}\rmd \theta_{D-1}^2~,
\end{equation}
with $\theta_i\in[-\frac{\pi}{2},~\frac{\pi}{2}]$ when $1<i<D-2$, and $\theta_{D-1}\in[0,~2\pi]$.

The Euclidean Einstein equations from (\ref{eq:onshell fundamental}) reduce to the constraint
\begin{equation}
    \qty(\frac{1}{N(\tau)}\frac{\rmd a}{\rmd \tau})^2 ={1-\frac{a^2}{L^2}-\frac{\kappa_D^2 Q^2 \,a^{-2(D-2)}}{(D-1)(D-2)}}\label{eq:derivative tau}\,.
\end{equation}
where the parameter $Q$ is a Noether charge of (\ref{eq:onshell fundamental}) associated with constant shifts in the axion field $\chi$ ($H_{D-1}:=\star d\chi$), $\chi\rightarrow\chi+\zeta$ with $\zeta\in\mathbf{R}$. It can be seen from the roots in (\ref{eq:derivative tau}) that $a(\tau)$ will always reach the same minimum and maximum values, $a_{\rm min}$ and $a_{\rm max}$ respectively, for any choice of the gauge parameter $N(\tau)$. To study the global geometry, we will consider the gauge choice $N(\tau)=1$ from this point onward. 

Moreover, the axion charge cannot take arbitrary values; there is a bound on the maximal size for the wormholes preparing the state, denoted by ``\emph{Nariai wormhole}", which can be obtained by extremizing (\ref{eq:derivative r, tau}) with respect to the scale factor $a$,
\begin{align}
    \kappa_D^2Q^2_{\rm max}&=L^{2(D-2)}(D-2)\qty(\frac{D-2}{D-1})^{D-2}\,.\label{eq:max size WH}
\end{align}
In this limit, the scale factor is a constant, given by
\begin{equation}
\begin{aligned}
    a_N:=a_{\text{max}}(Q_\text{max})&=a_{\text{min}}(Q_\text{max})=\sqrt{\frac{D-2}{D-1}}L\,.\label{eq:Conf scaling factor Nariai}
\end{aligned}
\end{equation}
\subsection{Two-copy Hartle Hawking state}
\begin{figure}[t!]
    \centering
    \includegraphics[width=0.7\textwidth]{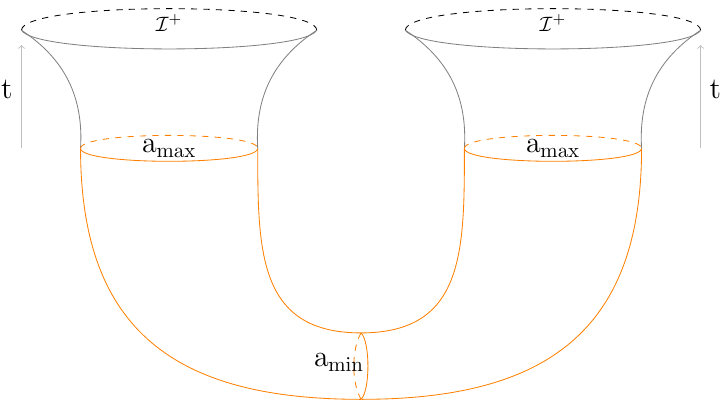}
    \caption{Preparation of the pair of axion-dS universes (gray surface) by slicing an Euclidean wormhole (orange surface) and performing the Wick rotation to global time $t$ in (\ref{eq:metric reg d}), which generates a two-copy HH state preparation. We illustrate the procedure by doing the analytic continuation at the maximum scale factor $a_{\rm max}$, while $a_{\rm min}$ represents the throat of the wormhole. This procedure results in expanding asymptotically dS universes}
    \label{fig:Setup}
\end{figure}

Our main interest is studying the real time evolution for these geometries, so we will now discuss the Lorentzian continuation, found by a simple Wick rotation $\tau\xrightarrow{}\rmi~ t$ in the global coordinate system (corresponding to the gauge choice $N(t)=1$):  
\begin{equation}
\rmd s^2=-\rmd t^2+a(t)^2\rmd\Omega_{D-1}^2\label{eq:metric reg d} ~,
\end{equation}
\begin{equation}
    \qty(\frac{\rmd a}{\rmd t})^2 ={-1+\frac{a^2}{L^2}+\frac{\kappa_D^2 Q^2 \,a^{-2(D-2)}}{(D-1)(D-2)}}\label{eq:derivative r, tau}\,.
\end{equation}
One can generate the Lorentzian geometry as a two-copy HH state preparation by slicing the wormhole at either $a=a_{\rm max}$ or $a=a_{\rm min}$, and performing the Wick rotation. The careful analysis of the scale factor in Lorentzian signature \cite{Aguilar-Gutierrez:2023ril} reveals that if one does this slicing at $a_{\rm max}$, we generate an expanding universe dominated by the cosmological constant term $\Lambda$; while the slicing for $a_{\rm min}$ leads to a contracting branch, due to high density of axion matter. We will focus on the first choice, illustrated in Fig. \ref{fig:Setup}, so that the resulting universes have dS asymptotics, and allow for a late-time evolution of gravitational observables, as well as a notion of weak gravity near $\mathcal{I}^+$. Moreover, by studying the propagation of matter inhomogeneities, one can find that the arrow of time \cite{Hartle:2011rb} follows opposite directions between the two universes, as we have illustrated in Fig. \ref{fig:Setup}.

\begin{figure}[t!]
    \centering
    \includegraphics[height=0.33\textwidth]{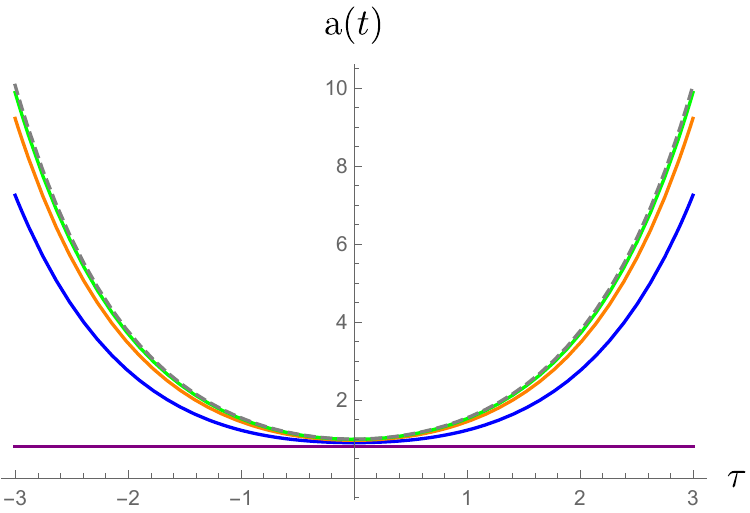}\hspace{0.1cm}\includegraphics[height=0.33\textwidth]{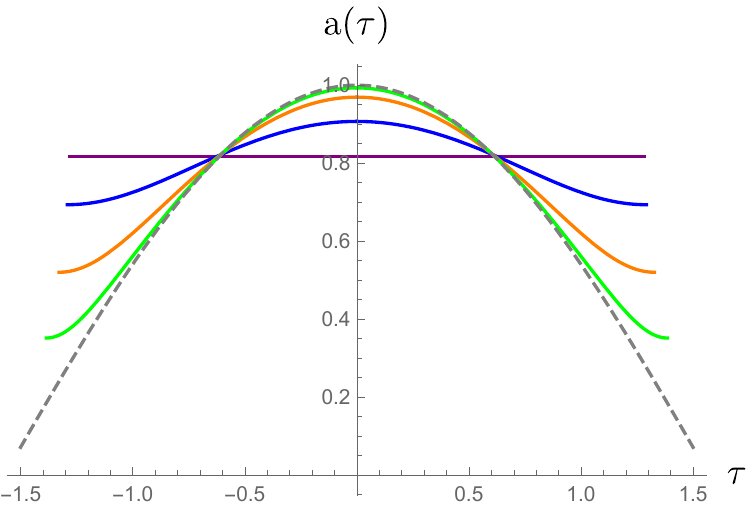}
    \caption{Scale factors in the regular global metric (\ref{eq:metric reg d}) with $D=4$ in Lorentzian (\textit{left}) and Euclidean (\textit{right}) signature (where $\tau=\rmi t$) for different values of the parameter $\mu:=Q/Q_{\rm max}$; $\mu = 1$ (purple), $0.9$ (blue), $0.6$ (orange) and $0.3$ (green), and $0$ (dashed gray) corresponding to pure dS space. All curves correspond to $L=1$.}
    \label{fig:scale-factors Lorentzian}
\end{figure}

In most of this section, we will take $a(t)$ in (\ref{eq:derivative r, tau}). For general values of $Q$, there are no analytic expressions for the scaling factors in this gauge. Numerical results in $D=4$ are shown in Fig. \ref{fig:scale-factors Lorentzian}.

For technical convenience, we will derive analytic expression using $D=3$ in (\ref{eq:derivative r, tau}) where there are analytic solutions for $a(\tau)$ in (\ref{eq:metric reg d}):
\begin{align}\label{eq:metric in r}
    \rmd s^2&=-\rmd t^2+\frac{L^2}{2}\qty(1+\cosh(\frac{2t}{L})\sqrt{1-\mu^2})\rmd\Omega_2^2~,\\
    \mu&:=\frac{Q}{Q_{\rm max}}~.\label{eq: parametrization charge}
\end{align}
Identifying the allowed range to cover the entire geometry, one sees that in this coordinate system $\tau\sim\tau+\pi/L$. 

\subsection{Geodesics}
Lastly, we would like to identify geodesic trajectories in this background geometry. It can be shown that for each of the axion-dS universes, there is a unique geodesic passing through the wormhole at $\theta_i(\tau=\tau_{\rm max})=0$, corresponding to \cite{Aguilar-Gutierrez:2023hls}
\begin{equation}\label{eq:geodesic path}
\theta_i^{\rm (g)}(t)=0,    \quad 1\leq i\leq D-1,\quad t\geq0~.
\end{equation}
This solution is useful to describe the worldline a pair of the geodesic observers.

\section{Maximal volume}\label{sec:CV}
We consider maximal codimension-one volume surfaces that are anchored to the pain of observers, i.e.
\begin{equation}\label{eq:CV pro}
    \mathcal{C}_V(\Sigma)=\max_{\Sigma=\partial \mathcal{B}}{V}(\mathcal{B})~,
\end{equation}
where $\Sigma$ is a time slice, and $\mathcal{B}$ a bulk codimension-one hypersurface, whose boundary $\partial\mathcal{B}$ anchored at $\Sigma$, with a corresponding volume ${V}(\mathcal{B})$.

\begin{figure}[t!]
    \centering
    \includegraphics[width=0.75\textwidth]{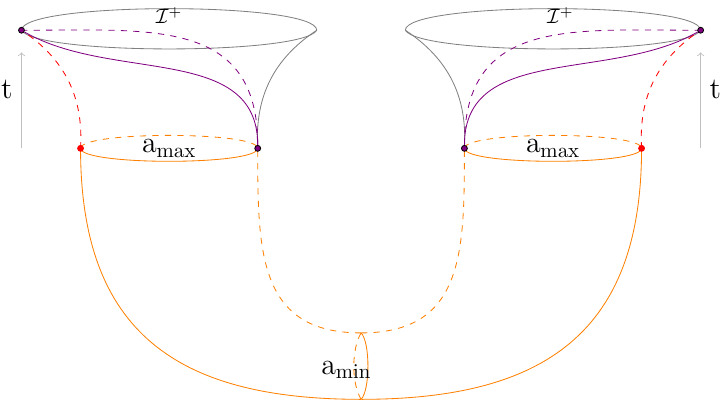}
    \caption{Maximal codimension-one volume surfaces (purple surfaces) anchored to pairs of worldline observer following the geodesic $\theta_i^{\rm (g)}(t)=0$ (red dashed lines) in axion-dS universes connected through an Euclidean axion wormhole (orange surface).}
    \label{fig:Csurface_adS}
\end{figure}

While we look for codimension-1 surfaces anchored to the geodesics in (\ref{eq:geodesic path}), given the symmetries of the problem, the spatial dependence on the extremal surface will only depend on one of the polar angles, $\theta_1$ in (\ref{eq:metric regular}). For this reason, from now on, we will denote $\theta_1\equiv\theta$ for notational convenience. The setting is illustrated in Fig. \ref{fig:Csurface_adS}.

Using (\ref{eq:metric reg d}) in (\ref{eq:CV pro}) leads to
\begin{equation}\label{eq:most general parametrization}
\begin{aligned}
    \mathcal{C}_V=\int\rmd\Omega_{D-2}\int\rmd\lambda&\sqrt{-\qty({t'(\lambda)})^2+a(t)^2\qty({\theta'(\lambda)})^2}~a(t)^{D-2}\cos^{D-2}\theta~,
\end{aligned}
\end{equation}
where primes indicate differentiation. Notice that this functional does not include the volume of the Euclidean wormhole connecting the universes, since it is independent of the Lorentzian time.

Notice that there are no conserved charges for the functional above; so we need to solve for the extremal surfaces ($\theta(\lambda)$, $t(\lambda)$) described by a non-linear coupled system of second-order ordinary differential equations, given by
\begin{align}
    0=&a(t) \cos (\theta ) \left(a'(t) \theta '(\lambda ) \left((D-1) a(t)^2 \theta '(\lambda )^2-D t'(\lambda )^2\right)+a(t) \left(\theta '(\lambda
   ) t''(\lambda )-\theta ''(\lambda ) t'(\lambda )\right)\right)\nonumber\\
   &+(D-2) \sin (\theta ) t'(\lambda ) \left(t'(\lambda )^2-a(t)^2 \theta
   '(\lambda )^2\right)~,\\
   0=&a(t) \cos (\theta ) \left(\theta '(\lambda ) a'(t(\lambda )) \left((D-1) a(t)^2 \theta '(\lambda )^2-D t'(\lambda )^2\right)+a(t) \left(\theta
   '(\lambda ) t''(\lambda )-\theta ''(\lambda ) t'(\lambda )\right)\right)\nonumber\\
   &+(D-2) \sin (\theta ) t'(\lambda ) \left(t'(\lambda )^2-a(t)^2
   \theta '(\lambda )^2\right)~.
\end{align}
The evaluation also relies on the boundary conditions. First, the extremal surfaces are anchored to each of the worldline observers along $\theta=0$ in each of their universes, and since we evaluate the maximal volume connect the two observers through the Euclidean wormhole, the extremal surface also reaches the $t=0$ surface, which maximizes the volume for $\theta=\pi/2$ when the observer is located at $\theta=0$, as shown in Fig. \ref{fig:Csurface_adS}. Therefore, the boundary conditions can be expressed as:
\begin{equation}\label{eq:anchoring to clock observers}
    t(\theta=0)=t_{\rm obs}~,\quad t\qty[\theta=\frac{\pi}{2}]=0~,
\end{equation}
where $t_{\rm obs}$ is the physical time for the geodesic observers, for which we choose to synchronize their clocks for simplicity. As for the second equality in (\ref{eq:anchoring to clock observers}), this indicates that the surfaces anchored to the worldline observer reach the Euclidean wormhole (see Fig. \ref{fig:Csurface_adS}) coupling the two universes at $t(\theta)=0$, which is maximal for $\theta=\pi/2$ when the observers are located at $\theta=0$, given that $\theta\in[-\pi/2,~\pi/2]$ and the argument of the integral (\ref{eq:most general parametrization}) is positive definite over this range, as made manifest in the $\lambda=\theta$ gauge.

Also note that, given that the choice of boundary conditions (\ref{eq:anchoring to clock observers}) relies on the Euclidean wormhole coupling the pair of universes, the following evaluations does not apply for the pure dS space limit where the axion charge $Q=0$.

Therefore, we proceed using the $\lambda=\theta$ gauge where the boundary conditions (\ref{eq:anchoring to clock observers}) can be most conveniently implemented, such that (\ref{eq:most general parametrization}) becomes:
\begin{equation}\label{eq:CV theta parametrization}
\begin{aligned}
    \mathcal{C}_V&=\Omega_{D-2}\int_{0}^{\pi/2}\rmd\theta~\mathcal{L}_V~,\\
    \mathcal{L}_V&:=\sqrt{-\qty({t'({\theta})})^2+a(t)^2}~a(t(\theta))^{D-2}\cos^{D-2}\theta~.
\end{aligned}
\end{equation}
The equations of motion (EOM) for the extremal surface from $\mathcal{C}_V$ can be expressed as:
\begin{equation}\label{eq:EOM t(theta)}
\begin{aligned}
    0=&\cos (\theta ) a(t(\theta )) \left(a'(t(\theta )) \left((D-1) a(t(\theta ))^2-D~t'(\theta )^2\right)+a(t(\theta )) t''(\theta )\right)\\
    &+(D-2)\sin (\theta ) t'(\theta ) \left(t'(\theta )^2-a(t(\theta ))^2\right)~.
\end{aligned}
\end{equation}
Lastly, we can explicitly check that the boundary conditions (\ref{eq:anchoring to clock observers}) are compatible with the extremization of the functional $\mathcal{C}_{\rm CMC}$ (\ref{eq:regions CMC}). First notice that for the total variation $\delta \mathcal{C}_V$ to vanish after imposing the EOM in (\ref{eq: CCMC adS}) requires that:
\begin{equation}\label{eq:total variation to vanish}
    \eval{\delta t(\theta)a(t(\theta))^{D-2}\cos^{D-2}\theta\tfrac{t'(\theta)}{\sqrt{-t'(\theta)^2+a(t(\theta))^2}}}_{\theta=0}^{\theta=\pi/2}=0~.
\end{equation}
This condition is satisfied by (\ref{eq:anchoring to clock observers}) since $t$ is fixed at $\theta=0$, while the term at $\theta=\pi/2$ necessarily vanishes for $D>2$.

Next, we will explicitly study the evolution of the volume. Since $a(t)$ in (\ref{eq:metric reg d}) is only known analytically for $D=3$ with arbitrary $Q\leq Q_{\rm max}$, and when $Q=Q_{\rm max}$ in arbitrary $D$, we will focus in these cases. However, the implementation is valid for more arbitrary cases.

\subsection{\texorpdfstring{$D=3$}{} case}\label{sec:CV 3D}
\begin{figure}[t!]
    \centering
    \includegraphics[height=0.32\textwidth]{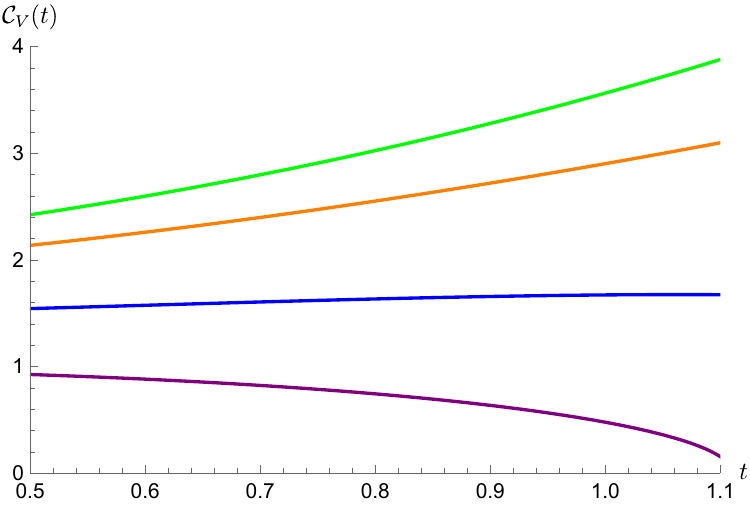}\hspace{0.1cm}\includegraphics[height=0.32\textwidth]{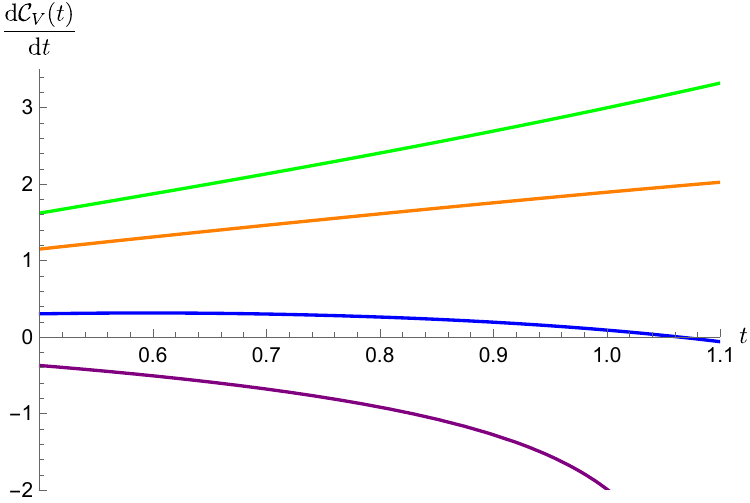}
    \caption{\textit{Left}: Maximal volume of the axion-dS universes in $D=3$, and \textit{Right}: its rate of growth for different values of $\mu = 1$ (purple), $0.9$ (blue), $0.6$ (orange) and $0.3$ (green). As the axion charge density decreases, the observables displays a tendency towards a smaller maximal volume. More details about the reason for the decrease are explain in the main text.}
    \label{fig:CV 3D}
\end{figure}

We now numerically solve the extremal surfaces $t(\theta)$ in (\ref{eq:EOM t(theta)}) with (\ref{eq:anchoring to clock observers}) for the $D=3$ axion-dS scale factor (\ref{eq:metric in r}) and substitute it in (\ref{eq:CV theta parametrization}), which results in Fig. \ref{fig:CV 3D}.
The range of the $t$ coordinate displayed in Fig. \ref{fig:CV 3D} is taken where the numerical solutions are found to be stable. {We are considering $t\sim\mathcal{O}(1)~L$ given that the numerical results become less reliable at late times. We confirmed (albeit in the presence of noise) that the late-time behavior follows a similar trend, the volume is nearly exponentially increasing for different charge ratios ($\mu=0.3$, $0.6$ in the plot).} The reason for this is encoded in the scale factors (\ref{eq:metric in r}), which are nearly exponentially increasing at late time. Since the extremal surfaces are anchored between the worldline observers $\theta=0$ and the wormhole $t(\theta=\pi/2)=0$, the time dependence encoded in the scale factor (\ref{eq:CV theta parametrization}) allows for the volumes to reach large late-time values. However, this issue is more subtle as there is also competition with the axion density. When the parameter $\mu\sim1$, the volume decreases to a vanishing value at a finite time scale, particularly in the Nariai limit ($\mu=1$). This situation is analyzed in more detail in the following subsection.

\subsection{\texorpdfstring{$Q=Q_{\rm max}$}{} case}\label{sec:CV Nariai}
\begin{figure}[t!]
    \centering
    \includegraphics[height=0.33\textwidth]{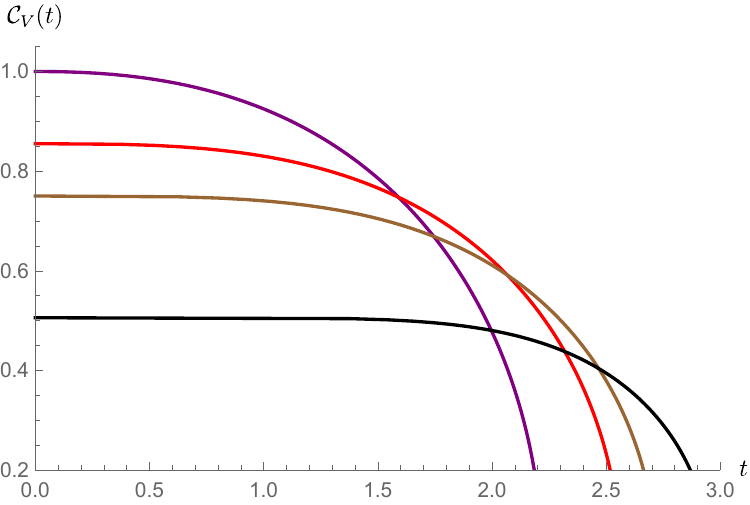}\hspace{0.1cm}\includegraphics[height=0.33\textwidth]{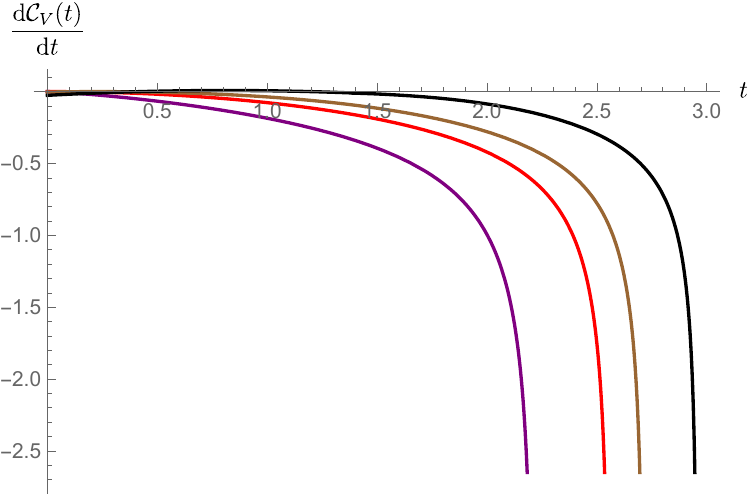}
    \caption{\textit{Left}: Maximal volume growth of the axion-dS universes in the Nariai limit ($\mu=1$, corresponding to a pair of Einstein static universes) according to the maximal volume, and \textit{Right}: its rate of growth for $D = 3$ (purple), $4$ (red), $5$ (brown) and $10$ (black). The maximal volume decreases over time.}
    \label{fig:CV Nariai}
\end{figure}

As we saw in (\ref{eq:Conf scaling factor Nariai}), the scale factor is a constant $a_N=\sqrt{\frac{D-2}{D-1}}L$. We can then write the conserved charge for (\ref{eq:CV theta parametrization}) as
\begin{equation}
\begin{aligned}
    P&=\dv{\mathcal{L}_V}{t'(\lambda)}\\
    &=-\frac{a_N^{D-2}\cos^{D-2}\theta ~t'({\lambda})}{\sqrt{-\qty(t'({\lambda}))^2+a(t)^2\qty(\theta'({\lambda}))^2}}~.\label{eq:momentum}
\end{aligned}
\end{equation}
One can then find integral expressions for:
\begin{align}
    \mathcal{C}_V&=\int_{0}^{\pi/2}\frac{a_N^{2D-3}\cos^{2(D-2)}\theta~\rmd\theta}{\sqrt{P^2+a_N^{2(D-2)}\cos^{2(D-2)}\theta}}~,\label{eq:CV Nariai}\\
    t_{\rm obs}&=-\int_{0}^{\pi/2}\frac{P~a_N~\rmd\theta}{\sqrt{P^2+a_N^{2(D-2)}\cos^{2(D-2)}\theta}}~.\label{eq:t Nariai}
\end{align}
The cases $D=3$, $4$, $5$ and $10$ are plotted in Fig. \ref{fig:CV Nariai}.

As we saw in Sec. \ref{sec:CV 3D}, the volume decreases until it reaches a vanishing value. However, the analysis of integrals above (\ref{eq:CV Nariai}) and (\ref{eq:t Nariai}) shows its evolution cannot take place indefinitely, instead $t_{\rm obs}$ can be at most $t_{\rm crit}:=\frac{\pi}{2}a_N$, which occurs when $\mathcal{C}_V$ reaches $0$ for $P$ large enough. The reason for this is related to the scale factor being constant. There are no longer extremal surfaces obeying the boundary conditions (\ref{eq:anchoring to clock observers}) once we reach the critical time $t_{\rm crit}$, corresponding to the limit where the extremal surfaces become null-like; and as a result, there is no longer an appropriate measure of codimension-one volume. We will explore how this situation may be modified in Sec. \ref{sec:CAny Nariai}.

We remark that the Nariai limit is not asymptotically dS, as one notices by the fact that the scale factor is a constant; its Lorentzian interpretation is that of a pair of Einstein static universes that are coupled to one another through the Euclidean wormhole. Nevertheless, the results presented in this section are meant to provide a better analytic understanding of the late-time behavior of the axion-dS universes where $\mu\sim 1$ displayed in Fig. \ref{fig:CV 3D}.

\section{Bulk curvature invariants in constant mean curvature slices}\label{sec:CAny}
\begin{figure}[t!]
    \centering
    \includegraphics[width=\textwidth]{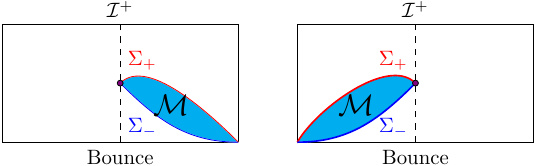}
\caption{Proposal for evaluating the volume of CMC slices for axion-dS universes connected through a quantum bounce (explained in Sec. \ref{sec:intro}). We anchor codimension-one extremal surfaces $\Sigma_-$ and $\Sigma_+$ (in blue and red respectively) to worldline geodesic observers (represented by black dashed lines), which determine the past and future boundaries of the spacetime region $\mathcal{M}$, where $\mathcal{C}^{\epsilon}$ \eqref{eq:Volepsilon} is evaluated. The evolution of these observables is defined with respect to the global time, or alternative by the observer's proper time (syncronnized in the past). The precise profile of the $\Sigma_\pm$ slices is determined by the extremization of eq.~\eqref{eq:regions CMC}.}
    \label{fig:ProposaladS}
\end{figure}

We are interested in codimension-one observables generalizing the maximal volume in the previous section.
First, we define the set of codimension-one observables defined in \cite{Belin:2021bga,Belin:2022xmt,Jorstad:2023kmq}
\begin{equation}\label{eq:Volepsilon}
    \mathcal{C}^\epsilon \equiv\int_{\Sigma_\epsilon}d^{D-1}\sigma\,\sqrt{h}~F[g_{\mu\nu},\,\mathcal{R}_{\mu\nu\rho\sigma},\,\nabla_\mu]~,
\end{equation}
where $F[g_{\mu\nu},\,\mathcal{R}_{\mu\nu\rho\sigma},\,\nabla_\mu]$ is an arbitrary scalar functional of $D$-dimensional curvature invariants of the bulk region $\mathcal{M}$, which is covariantly defined by extremizing a combination of codimension-one and codimension-zero volumes with different weights, given by
\begin{equation}\label{eq:regions CMC}
\begin{aligned}
    \mathcal{C}_{\rm CMC}&=\alpha_+\int_{\Sigma_+}\rmd^{D-1}\sigma\,\sqrt{h}+\alpha_-\int_{\Sigma_-}\rmd^{D-1}\sigma\,\sqrt{h}+ \frac{\alpha_B}{L}\int_{\mathcal{M}}\rmd^{D}x\sqrt{-g}~,
\end{aligned}
\end{equation}
where $\Sigma_{\pm}$ are the future (past) boundaries of $\mathcal{M}$ anchored at the locations in (\ref{eq:anchoring to clock observers}), such that $\partial\mathcal{M}=\Sigma_+\cup\Sigma_-$, as shown in Fig.~\ref{fig:ProposaladS}.\footnote{We will label with $\epsilon=+,-$ the quantities defined on the codimension-one surfaces $\Sigma_{\pm}$.  }

We denote $h$ as the determinant of the induced metric on $\Sigma_{\pm}$. 
The coefficients $\alpha_{\pm}$ and $\alpha_B$ are dimensionless positive constants, therefore the extremization of the functional \eqref{eq:regions CMC} defines CMC slices, for which the extrinsic curvature is given by \cite{MARSDEN1980109,Belin:2022xmt}
\begin{equation}\label{eq:Kepsilon}
    K_\epsilon:=\eval{K}_{\Sigma_\epsilon} = -\epsilon\frac{\alpha_B}{\alpha_\epsilon~L}~.
\end{equation}
Here we consider outward-pointing vectors with respect to the surfaces $\Sigma_\epsilon$ to be future-directed.

We now perform the explicit evaluation of (\ref{eq:Volepsilon}) using the background geometry (\ref{eq:metric reg d}, \ref{eq:derivative r, tau}) and the same boundary conditions in (\ref{eq:anchoring to clock observers}):
\begin{equation}\label{eq:CAny proposal}
\begin{aligned}
    \mathcal{C}^\epsilon=\Omega_{D-2}\int\rmd\theta~&\sqrt{-\qty({t
    }'({\theta}))^2+a(t)^2}~ b(t,~\theta)~a(t)^{D-2}\cos^{D-2}\theta~,
\end{aligned}
\end{equation}
where $b(t,~\theta)$ is an arbitrary function corresponding to the choice of the functional $F[\dots]$ in (\ref{eq:Volepsilon}) for the background (\ref{eq:metric reg d}).

Meanwhile, we can evaluate the contribution of the spacetime volume and codimension-one volumes of $\Sigma_\pm$ in (\ref{eq:regions CMC}) with (\ref{eq:metric reg d}, \ref{eq:derivative r, tau}). This leads to
\begin{equation}\label{eq: CCMC adS}
    \mathcal{C}_{\rm CMC}=\Omega_{D-1}\sum_\epsilon \alpha_\epsilon\int_{\Sigma_\epsilon}\rmd\theta\qty[\sqrt{-\qty(t'(\theta))^2+a(t)^2}~a(t)^{D-2}+K_\epsilon\int\rmd t~ a(t)^{D-1}]\cos^{D-2}\theta~,
\end{equation}
with $K_\epsilon$ given in (\ref{eq:Kepsilon}).

Next, we need to extremize (\ref{eq: CCMC adS}) to find the extremal surfaces $t(\theta)$. First, notice that the argument about the boundary conditions in (\ref{eq:anchoring to clock observers}) is still consistent with $\delta\mathcal{C}_{\rm CMC}=0$ in (\ref{eq:regions CMC}) after imposing EOM (\ref{eq:EOM CAny CMC}) and (\ref{eq:total variation to vanish}). This means that the parameter $K_{\epsilon}$ does not change the argument in Sec. \ref{sec:CV}.

Notice that there is no longer a conserved momentum associated with the global time $t(\theta)$, instead the EOM for (\ref{eq: CCMC adS}) read:
\begin{equation}\label{eq:EOM CAny CMC}
\begin{aligned}
   0=&\cos (\theta ) a(t(\theta ))^3 \left((D-1) a'(t(\theta ))+K_{\epsilon } \sqrt{a(t(\theta ))^2-t'(\theta )^2}\right)\\
   &-\cos (\theta ) a(t(\theta
   )) t'(\theta )^2 \left(D~ a'(t(\theta ))+K_{\epsilon } \sqrt{a(t(\theta ))^2-t'(\theta )^2}\right)\\
   &+a(t(\theta ))^2 \left((2-D) \sin (\theta
   ) t'(\theta )+\cos (\theta ) t''(\theta )\right)+(D-2) \sin (\theta ) t'(\theta )^3
\end{aligned}
\end{equation}
where the boundary conditions are the same as those in (\ref{eq:anchoring to clock observers}) as in Fig. \ref{fig:Csurface_adS}.

A technical issue in the evaluation of (\ref{eq:EOM CAny CMC}) is that, as we have explained in Sec. \ref{sec:review}, the scale factor in the global metric needs to be determined by numerical methods for general charge ratio $\mu:=Q/Q_{\rm max}$ in higher dimensions. 

As discussed in Sec. \ref{sec:CV}, the scale factor $a(t)$ in the regular global metric (\ref{fig:scale-factors Lorentzian}) generically needs to be found numerically and plug back in (\ref{eq:EOM CAny CMC}). To properly illustrate the procedure, we will work with the exactly solvable cases for $a(t)$, namely $D=3$ with $0\leq\mu\leq1$ (\ref{eq:metric in r}), and for $\mu=1$ with $D\geq3$ (\ref{eq:Conf scaling factor Nariai}). We expect similar results to hold $D=4$ and arbitrary $0<\mu\leq1$.

Our analysis for the relational observables will now be focused on the volume of the CMC slices, that is when $b(t,~\theta)=1$ in (\ref{eq:CAny proposal}). However, since $b(t,~\theta)$ in (\ref{eq:CAny proposal}) is an arbitrary function, it can modify significantly the behavior of $\mathcal{C}^\epsilon$ for other choices. Nevertheless, given that $b(t,~\theta)$ does not enter the EOM for the extremal surfaces, it would not change the fact that the extremal surfaces do no longer exist after a critical time $t_{\rm crit}$ which we explain Sec. \ref{sec:CAny Nariai}.

\subsection{\texorpdfstring{$D=3$}{} case}
\begin{figure}[t!]
    \centering
    \includegraphics[height=0.33\textwidth]{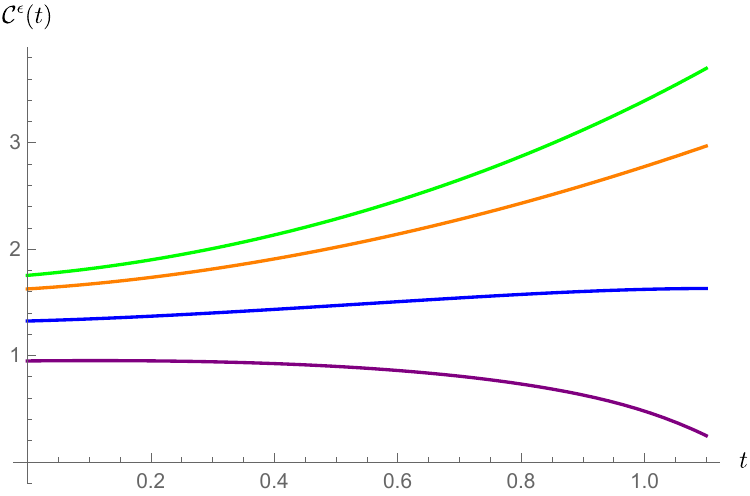}\hspace{0.1cm}\includegraphics[height=0.33\textwidth]{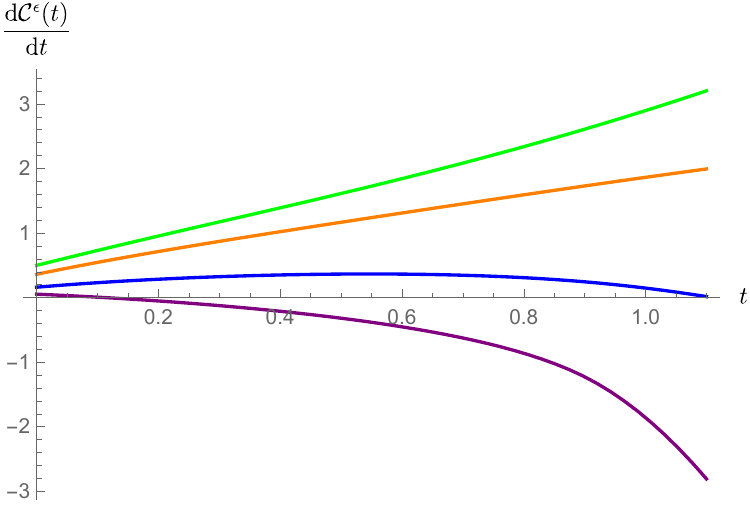}
    \caption{Curvature invariants evaluated on CMC slices (\ref{eq: CCMC adS}) (\textit{left}) and the corresponding rate of growth (\textit{right}). The parameters are the same as displayed in Fig. \ref{fig:CV 3D}, with the addition of $K_-=1/L$ and $F[\dots]=1$, corresponding to the volume of the CMC slices. The rate of growth is increased with respect to that in the maximal volume proposal in Fig. \ref{fig:CV 3D} (corresponding to $K_\pm=0$).}
    \label{fig:CAny3D}
\end{figure}

We will perform a numerical analysis similar to that in Sec. \ref{sec:CV 3D}. Using the scale factor in (\ref{eq:metric in r}) in the EOM (\ref{eq:EOM CAny CMC}) subject to the boundary conditions (\ref{eq:anchoring to clock observers}) leads to the $t(\theta)$ surfaces. The results for $F[\dots]=1$ (\ref{eq:Volepsilon}) (corresponding to $b(t,~\theta)=1$ in (\ref{eq:CAny proposal})) are displayed in Fig. \ref{fig:CAny3D}.

The evolution of the class of observables in (\ref{eq:CAny proposal}) is very similar to the findings in Sec. \ref{fig:CV 3D} for the maximal volume, for which $K_{\epsilon}=0$. The main difference is in the rate of growth that increases given that the CMC slices tilt towards the past as $K_->0$ is increased, which means that they increase in size as the worldline observer moves towards the future in asymptotically dS space. This allows for an enhanced rate of growth of codimension-one volumes evaluated on the CMC slices. In contrast, we would recover a decrease in the rate of growth with respect to the maximal volume when using $K_+$, which is negative as seen in (\ref{eq:Kepsilon}). 

\subsection{\texorpdfstring{$Q=Q_{\rm max}$}{} case}\label{sec:CAny Nariai}
\begin{figure}[t!]
    \centering
    \includegraphics[height=0.33\textwidth]{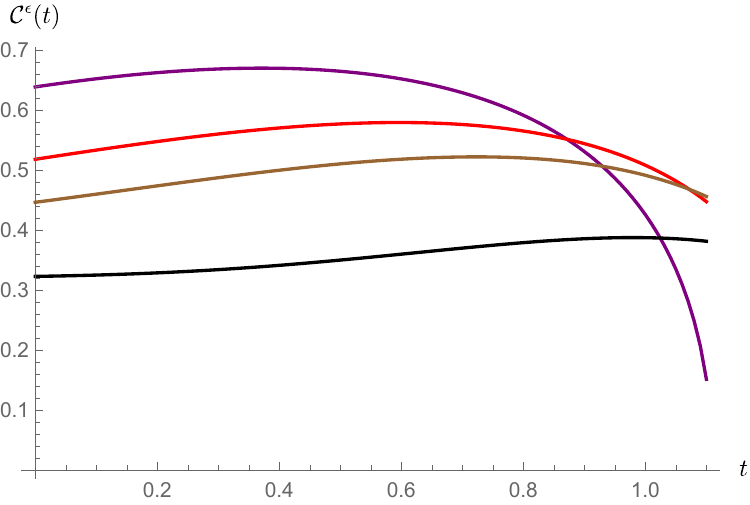}\hspace{0.1cm}\includegraphics[height=0.33\textwidth]{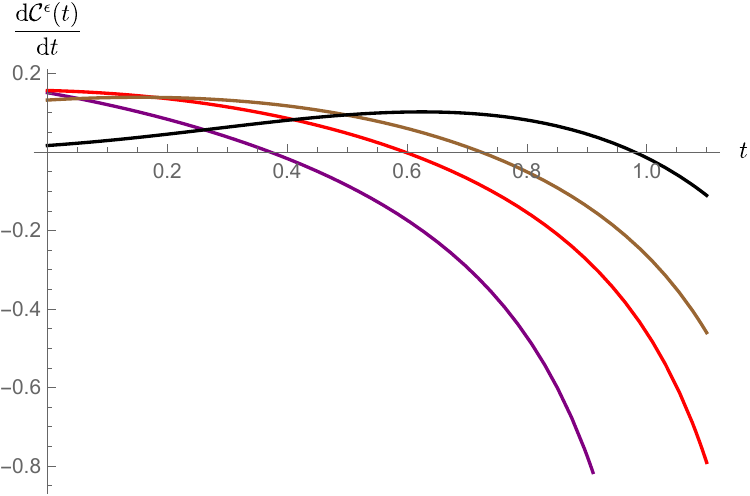}\\
    \includegraphics[height=0.33\textwidth]{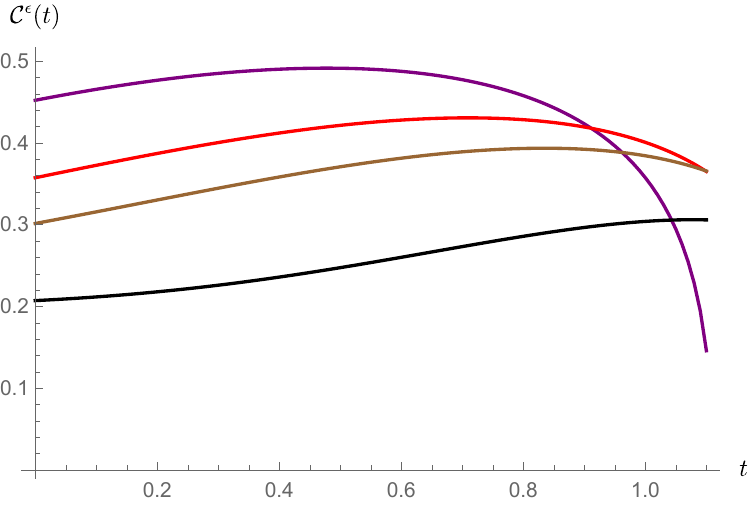}\hspace{0.1cm}\includegraphics[height=0.33\textwidth]{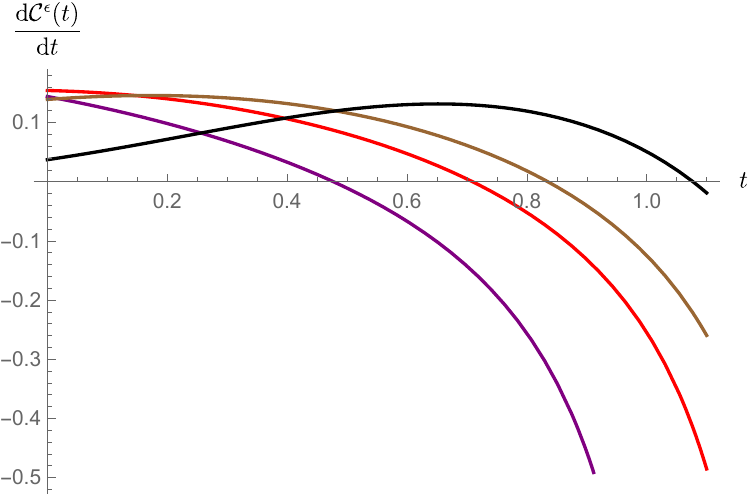}
    \caption{Curvature invariants evaluated on CMC slices (\ref{eq: CCMC adS}) (\textit{left}) and the rate of growth (\textit{right}) with $K_-=5/L$ (above), and $K_-=10/L$ (below). We employed the same numerical parameters as in Fig. \ref{fig:CV Nariai}, and $F[\dots]=1$. We notice that as $K_-$ increases, the rate of growth of $C^\epsilon$ at given $t$ is also enhanced.}
    \label{fig:CAnyN}
\end{figure}

First, notice that although in the $Q=Q_{\rm max}$ regime $a(t)=a_N$ is a constant, there is still $t(\theta)$ dependence in the functional (\ref{eq: CCMC adS}), and as a result, there are no conserved charges for $\mathcal{C}_{\rm CMC}$, unlike in Sec. \ref{sec:CV Nariai}. Thus, we will proceed with the numerical analysis as in the previous subsection. We solve the EOM (\ref{eq:EOM CAny CMC}) with the scale factor (\ref{eq:Conf scaling factor Nariai}) and the boundary conditions (\ref{eq:anchoring to clock observers}). The resulting evolution for the CMC observables is displayed in Fig. \ref{fig:CAnyN}.

The results are reminiscent of those we found in Sec. \ref{sec:CV Nariai}. Namely, there is a critical time, $t_{\rm crit}$, for which the extremal surface no longer exists for the set of boundary conditions in (\ref{eq:anchoring to clock observers}). However, we see that by increasing the value of $K_{\epsilon}$ ($\epsilon=-$ in Fig. \ref{fig:CAnyN}), it takes longer for $C^\epsilon$ to decrease, indicating that the $t_{\rm crit}$ can be prolonged by properly increasing $K_\epsilon$. It would be interesting to find if there is an asymptotic regime where $t_{\rm crit}\rightarrow\infty$.

\section{Discussion}\label{sec:Discussion}
To summarize, we investigated the evolution of the simplest relational observables in axion-dS universes, including codimension-one volumes and more general curvature invariant functionals. These describe spatially closed bouncing FLRW cosmologies with an ultra-stiff fluid (with an EOM $\rho=p$) that are coupled through a Euclidean wormhole. We focused on the case where these cosmologies have dS asymptotics. A worldline geodesic observer resides in each universe. Our results show that generically these relational observables evolve with a nearly exponential dependence on the observer's time if the axion charge is low enough and that the evolution can instead decrease when the axion charge approaches the maximal value allowed by dS space, which is referred to as the Nariai limit. The reason for this is that the extremal surface may become null-like at late enough times depending on the axion charge density supported on the dS universes. The main results are shown in Figs. \ref{fig:CV 3D}, \ref{fig:CV Nariai}, \ref{fig:CAny3D}, \ref{fig:CAnyN}.

There were certain limitations in the numerical part of the work which we comment on below. First, the analysis of the time dependence was performed for $t\sim\mathcal{O}(1)~L$ given the limitations in the numerical integration at late times, where more noise is introduced. Nevertheless, we observed (up to numerical noise) that they still follow an exponentially increasing behavior. 

Second, as we mentioned in the main text, we focused on solving $D=3$ for any value $Q$; and the $Q=Q_{\rm max}$ limit for $D\geq3$ for technical convenience. In contrast, the numerical methods for solving (\ref{eq:EOM CAny CMC}) need further improvement elaboration when $a(t)$ in the global coordinate system (\ref{eq:metric reg d}) can only be solved numerically. Given that the scale factors in $D=4$ do not change significantly with respect to the $D=3$ case, we suspect our results for the time evolution of codimension-one volumes would be very similar in that case.

All in all, we hope this study provides new insights into relational observables in closed universes with more intricate topology.

\subsection{Outlook}We now comment on future avenues for investigation.

First, our work assumes that the dynamical reference frames where we anchor the relational observables are classical, based on the formalism in \cite{Goeller:2022rsx}. However, the formalism can be naturally extended in the framework of quantum reference frames. To take steps towards a quantum description of the system that we studied, one could work with a simplified version of the model through dimensional reduction, as explained in \cite{Aguilar-Gutierrez:2023hls}, to a two-dimensional dilaton gravity version of our model. One could then quantize the axionic matter content, following similar considerations as \cite{Kolchmeyer:2024fly}, where the observers introduced in this work can be represented through matter particles with finite masses. The type of geometric observables that we explored in this work might be less well defined when the location of the pair of observers is not fixed along a classical path; however, one could in principle derive a more general gauge invariant operator algebra, generalizing the classical, as defined in \cite{DeVuyst:2024pop,DeVuyst:2024uvd} for more general background spacetimes\footnote{See also \cite{Blommaert:2025bgd} for a discussion on the algebra of observables in the background employed in this work, albeit still in the semiclassical approximation.} that would generalize the classical observables in this work. Another important generalization would be to introduce multiple observers within each universe; this might help us to study subsystem relativity (e.g. \cite{Hoehn:2023ehz}) within each of the universes. More generally, it would be interesting to consider how higher genus bulk topology might modify the relational atlas introduced in \cite{Goeller:2022rsx} to cover the spacetime regions that can be accessed by the union of the multiple reference frames.

Second, the observables in our study are conjectured to represent holographic complexity in holographic settings \cite{Belin:2021bga,Belin:2022xmt}, which we have directly connected to relational observables according to the dynamical reference frames. While we have focused on a very specific bulk theory, we expect that similar conclusions can be found in more general gravity theories with dynamical reference frames, which is worth to inspect in closer detail. To properly assign a holographic complexity interpretation, one needs to further relate our works with that literature. For instance, we might considering the circuit complexity of an explicit holographic dual theory for each of the universes that are maximally entangled with its copy. However, according to Nielsen's geometric approach to circuit complexity \cite{Nielsen2005,nielsen2006quantum,dowling2008geometry} complexity growth should not be greater than linear \cite{Aguilar-Gutierrez:2023nyk}. This might imply that our findings represent something closer to Krylov operator complexity \cite{Parker:2018yvk}. It would be interesting to develop a microscopic interpretation of the observables in this work in terms of a dual quantum theory, conjectured to be located near $\mathcal{I}^+$ \cite{Aguilar-Gutierrez:2023hls}.

Third, we have focused on codimension-one observables in this work; however, there is a larger class of codimension-zero curvature-invariant observables to generalize our results \cite{Belin:2022xmt}. For instance, one could study how generic are our finding on the increase and decrease of extremal volumes depending on the axion charge and the cosmological constant within the larger class of observables in \cite{Belin:2022xmt}. Another natural extension of the results would be to extend our results regarding maximal volumes in asymptotically flat or AdS axion wormhole spacetimes. It has been noted that the holographic dictionary shows new puzzles once Euclidean wormholes are attached to disconnected universes \cite{Hertog:2017owm,Loges:2023ypl}. It would be interesting to investigate what modifications this would lead to for the observables considered in this work. For instance, previous work on the maximal volume for asymptotically AdS wormholes in $D=3$ gravity can be found in \cite{Zolfi:2023bdp}.

Lastly, we have only investigated the time evolution of different relational observables in the background geometry without perturbations. However, asymptotically dS spacetimes are expected to display the switchback effect \cite{Aguilar-Gutierrez:2023pnn,Baiguera:2023tpt,Baiguera:2024xju}, which describes a decrease in the growth of the class of curvature invariants in \cite{Belin:2021bga,Belin:2022xmt} due to the appearance of shockwave perturbations in the geometry, motivated by epidemic growth of perturbations in quantum circuit dual models. It might be interesting to study this case by sending radial energy pulses from the perspective of the worldline observers and accounting for the backreaction in the geometry (\ref{eq:metric reg d}).

\section*{Acknowledgements}
I thank Stefano Baiguera, Mehrdad Mirbabayi, and Nicolò Zenoni for useful discussions and the HECAP group at the International Centre for Theoretical Physics for their hospitality and support during part of the project. During earlier stages of this project, the work of SEAG was partially supported by the FWO Research Project G0H9318N and the inter-university project iBOF/21/084 during the start of this work. SEAG is now supported by the Okinawa Institute of Science and Technology Graduate University. This project/publication was also made possible through the support of the ID\#62312 grant from the John Templeton Foundation, as part of the ‘The Quantum Information Structure of Spacetime’ Project (QISS), as well as Grant ID\# 62423 from the John Templeton Foundation. The opinions expressed in this project/publication are those of the author(s) and do not necessarily reflect the views of the John Templeton Foundation.

\appendix
\section{Notation}\label{app:notation}
\paragraph{Acronyms}
\begin{itemize}[noitemsep]
\item (A)dS: (anti-)de Sitter.
    \item CFT: conformal field theory.
\item CMC: Constant mean curvature.
\item EOM: Equations of motion.
\item HH: Hatle-Hawking.
\end{itemize}
\paragraph{Symbols}
\begin{itemize}
    \item $\mathcal{C}^\epsilon$ (\ref{eq:CV pro}): Codimension-one curvature-invariant observables.
    \item $\mathcal{C}_V$ (\ref{eq:CV pro}): Maximal volume.
    \item $t_{\rm obs}$ \eqref{eq:t Nariai}: Observer's time in global coordinates \eqref{eq:metric regular}, which is bounded by the critical time $t_{\rm crit}$.
    \item $F[g_{\mu\nu},\,\mathcal{R}_{\mu\nu\rho\sigma},\,\nabla_\mu]$: Arbitrary scalar functional of $D$-dimensional curvature invariants of the bulk region $\mathcal{M}$.
    \item $\Sigma_\pm$: Past and future-like boundaries of the CMC slices, where $\partial\mathcal{M}=\Sigma_+\cup\Sigma_-$.
     \item $\mathcal{L}_V$ (\ref{eq:CV theta parametrization}): Maximal volume Lagrangian.
    \item $t(\lambda)$, $\theta(\lambda)$: Coordinate of the extremal surfaces in terms of a general parametrization $\lambda$.
    \item $K_\pm$ (\ref{eq:Kepsilon}): Corresponding curvature at $\Sigma_\pm$
    \item $\mathcal{B}$: codimension-one hypersurface.
    \item $\mathcal{C}_{\rm CMC}$ (\ref{eq: CCMC adS}): Functional to determine the evolution of the CMC slices.
    \item $a(\tau)$ and $N(\tau)$: Scale factor and the lapse in \eqref{eq:metric regular}.
    \item $a_N$ \eqref{eq:Conf scaling factor Nariai}: Maximal scale factor
    \item $Q$: Axion charge density.
    \item $L$ (\ref{eq:Lambda C.C.}): dS length scale in the cosmological constant $\Lambda$.
    \item $\rmd\Omega_{D-1}$ (\ref{eq:metric to be gauged}): Infinitesimal $(D-1)$-sphere volume element.
    \item $P$ (\ref{eq:momentum}): Conserved charge related to time shifts.
    \item $Q_{\rm max}$ \eqref{eq:max size WH}: Maximal axionic charge (for Nariai universes).
    \item $\mu:=\frac{Q}{Q_{\rm max}}$ dimensionless ratio with respect to the maximal axion charge density \ref{eq: parametrization charge}.
\end{itemize}

\bibliographystyle{JHEP}
\bibliography{references.bib}
\end{document}